# ПАРАМЕТРИЧЕСКИЙ СИНТЕЗ МУЛЬТИСЕРВИСНЫХ ТЕЛЕКОММУНИКАЦИОННЫХ СИСТЕМ С ПРИМЕНЕНИЕМ МАТЕМАТИЧЕСКОЙ МОДЕЛИ МНОГОСЛОЙНОГО ГРАФА

АГЕЕВ Д.В., ХАЙДАРА АБДАЛЛА

Харьковский национальный университет радиоелектроники

# MULTISERVICE TELECOMMUNICATION SYSTEMS PARAMETRICAL SYNTHESIS BY USING OF MULTILAYER GRAPH MATHEMATICAL MODEL

AGEYEV D.V., HAIDARA ABDALLA

Kharkiv national university of radioelectronics

***Аннотация.*** Проведенное исследование посвящено задаче параметрического синтеза мультисервисных телекоммуникационных систем. К основным свойствам телекоммуникационных систем, на которые обращено внимание в статье, отнесено многослойная структура образованная наложенными сетями и наличие у передаваемых потоков эффекта самоподобия. Для учета данных особенностей современных телекоммуникационных систем предложено использовать при описании структуры проектируемой системы многослойный граф, а для моделирования потоков в сети модели самоподобных процессов. Решение задачи параметрического синтеза сведена к задаче нелинейного программирования, которая решается с использованием метода градиентного спуска.

***Abstract.*** *This study is devoted to the problem of parametric synthesis of multi-service telecommunication systems. The main characteristics of telecommunication systems, which are brought to account in an article, are multilayer structure formed by the overlayed networks and presence flows with self-similarity effect. For accounting these features of modern telecommunications systems is proposed to use a multi-layered graph for describing the system structure, and self-similar processes model for modeling flows in a network. Solution of parametric synthesis problem is reduced to a nonlinear programming problem which is solved by using gradient descent method.*

## ВВЕДЕНИЕ

Этап проектирования является важной составляющей создания любой технической системы и именно от принятых на этом этапе решений зависит эффективность будущей системы. Среди задач, решаемых на этапе проектирования, есть задача выбора оптимальных значений параметров структурных элементов системы (задача параметрического синтеза). При проектировании телекоммуникационных систем к основным параметрам относятся пропускные способности каналов связи, емкости буферов и другие. Эффективность решения данной задачи во многом зависит от адекватности и корректности выбора математических моделей, как структурных, так и функциональных. Используемые в настоящее время методы параметрического синтеза базируются на применении классических математических моделей потоков, которые хорошо себя зарекомендовали при проектировании сетей с коммутацией каналов, таких как телефонные сети. Современные исследования трафика, передаваемого в телекоммуникационных системах [1, 2], показывают, что его статистические характеристики отличаются от тех, которые приняты в классической теории телетрафика. Это приводит к тому, что использование традиционных методов расчета параметров телекоммуникационных систем и их вероятностно-временных характеристик, основанных на пуассоновских моделях и формулах Эрланга, дает неоправданно оптимистические результаты, приводящие к недооценке нагрузки.

Последние исследования свойств информационных потоков в мультисервисных телекоммуникационных системах показали, что использование моделей самоподобных процессов (самоподобных потоков) позволяет более точно описывать трафик, передаваемый в данных системах.

Среди публикаций в направлении исследования самоподобных потоков существует большая нехватка в исследованиях, которые посвящены задачам параметрического синтеза телекоммуникационных систем с обеспечением общесистемных параметров качества обслуживания, базирующихся на использовании моделей самоподобных процессов.

Другой особенностью современных мультисервисных телекоммуникационных систем является их многоуровневая структура, которая наблюдается во многих аспектах рассмотрения. К многоуровневым структурам телекоммуникационных систем можно отнести функциональную, образованную уровнями модели ВОС; иерархическую, территориально распределенную, образованную уровнями NGN. В данной статье обращено внимание на многослойную структуру современных телекоммуникационных систем образуемую наложенными сетями. Учет последней играет большую роль так как процессы протекающие на разных уровнях тесно связаны и оказывают сильное влияние друг на друга.

В данной статье, базируясь на известных ранее исследованиях самоподобных потоков и на многоуровневой структуре, образованной наложенными сетями, предложен метод выбора пропускных способностей каналов связи мультисервисной телекоммуникационной сети.

## 1. АНАЛИЗ ПУБЛИКАЦИЙ ПО ТЕМЕ РЕШАЕМОЙ ЗАДАЧИ

Существует серия работ посвященных исследованию степени влияния самоподобных информационных потоков на качество обслуживания в узлах сети. Данные работы можно разделить на две группы.

Первой группе принадлежат работы посвященные исследованию вероятностно временных характеристик обслуживания с использованием средств имитационного моделирования или за счет проведения натурных экспериментов.

Так в работе [3] приведен обзор моделей фрактальных точечных процессов таких как: фрактальный ON/OFF источник, фрактальный дробовый точечный процесс, фрактальный биноминальный процесс. Данные модели позволяют создавать реализации самоподобных потоков в процессе имитационного моделирования.

Вторая группа работ посвящена исследованию качественных характеристик обслуживания с применением аналитических моделей.

В работе [4], используя модель фрактального броуновского движения в качестве модели самоподобного процесса, получено выражение для вероятности превышения длины очереди заданной пороговой величины для обслуживающих устройств с бесконечным объемом буфера. При введении дополнительных упрощений в работе получено выражение для требуемого объема буферного устройства, обеспечивающее заранее заданное качество обслуживания по вероятности потерь.

Используя результаты работы [4], в диссертационной работе [5] получено выражение для среднего времени задержки сообщения в узле и среднее время задержки сообщения в сети. Данные выражения использовались в работе [5] для проверки выполнения ограничения на максимально допустимую среднесетевую задержку сообщения в сети при решении задачи параметрического синтеза для случая передачи в сети потоков с параметром Херста, равном $H$=0,8. При этом для решения задачи выбора пропускной способности каналов связи использовался классический метод «квадратного корня» и не был предложен метод определения параметров потоков в каналах связи, получаемых при распределении потоков в сети, для случая самоподобных потоков.

В работе [6] используя модель фрактального броуновского движения получено выражения связывающее между собой такие параметры узла коммутации, как объем буферного устройства $N_б$, вероятность потери $P$ сообщения и интенсивность обслуживания µ с параметрами обслуживаемого самоподобного потока с параметрами λ - интенсивность поступления сообщений; $a$ - дисперсия процесса; $H$ – параметр Херста. В результате получено выражение для требуемой величины интенсивности обслуживания, при которой обеспечивается заданное ограничение на допустимую вероятность потерь сообщений

$$\mu = \lambda + \left(H^H(1-H)^{(1-H)}\sqrt{-2\ln(P)}\right)^{1/H} a^{1/(2H)} N_6^{-(1-H)/H} \lambda^{1/(2H)}. \tag{1}$$

Анализ данного выражения показывает, что при самоподобных потоках производительность системы растет медленнее при увеличении емкости буферных устройств по сравнению с случаем для пуассоновских потоков.

В работе [7] с применением средств имитационного моделирования проведено исследование формулы Норроса для пропускной способности системы и подтверждено, что она справедлива при широком диапазоне изменения входящих в нее параметров и рекомендуется инженерам для оценки пропускных способностей каналов связи. Однако приведенное выражение не позволяет учитывать временные затраты на передачу информационных потоков через сеть.

Современные телекоммуникационные системы являются большими сложными системами, которые тяжело поддаются математическому описанию. Помимо широко используемого выделения в структуре телекоммуникационной системы уровней иерархии по территориально-функциональному признаку (WAN, MAN, LAN), современные телекоммуникационные системы имеют технологическую многоуровневую структуру, которая образована наложенными сетями. Учету именно технологической многоуровневой структуры посвящена данная статья.

Известные ранее подходы при решении задач проектирования используют для учета многоуровневой природы современных телекоммуникационных систем последовательное решение задач проектирования для каждого из уровней отдельно. Результаты проектирования на одном из уровней являются исходными данными для остальных уровней сети. При этом в процессе проектирования не учитываются взаимосвязи и взаимозависимости между уровнями. В результате, итоговый вариант конфигурации сети не является оптимальным, а в ряде случаев может привести к нестабильной работе проектируемой сети при эксплуатации.

Для решения данной проблемы рядом авторов [8, 9] предлагается использовать математическую модель многоуровневой сети, которая построена с использованием упорядоченного набора графов. Топология каждого графа может отличаться, они могут иметь разный набор ребер, при этом, как правило, множество вершин графа с большим индексом (графа более высокого уровня) является подмножеством вершин графа с индексом на единицу меньшим (нижележащего уровня).

Описанная выше модель структуры многослойной сети имеет четкое соответствие вершин графа, описывающего каждый из уровней. В то же время большое количество систем имеет межуровневые связи более сложной природы, для которых приведенная выше модель теряет свою адекватность. Для устранения данного недостатка автором предлагается математическая модель структуры телекоммуникационной системы, базирующаяся на многослойном графе [10, 11]

## 2. ПОСТАНОВКА ЗАДАЧИ

Синтезируемая телекоммуникационная сеть (рис.1) обеспечивает передачу информационных потоков между узлами сети $a_i \in A$ являющимися источниками группового мультисервисного трафика и содержит кроме узлов-источников - транзитные узлы $z_i \in Z$, которые не продуцируют собственные информационные потоки и служат для передачи потоков других узлов сети.

<u>Исходными данными</u> для решения данной задачи является:

- данные об узлах сети $A = \{a_i\}$ – источниках информационных потоков;

- данные о перечне услуг $S = \{s_k\}$ предоставляемых в сети;

- данные о параметрах потоков $\gamma_{ij}^k$ передаваемых между всеми парами узлов $(a_i, a_j)$ отправитель-получатель при предоставлении услуги $s_k$;

- известна топология сети и маршруты передачи информационных потоков между всеми парами узлов отправитель-получатель;
- известны удельные затраты $\alpha_{ij}$ на организацию канала связи $(a_i, a_j)$ единичной пропускной способностью.

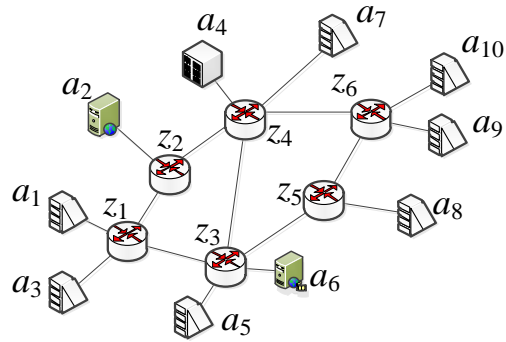

Рисунок 1. — Структура синтезируемой мультисервисной телекоммуникационной сети

В результате решения задачи параметрического синтеза <u>необходимо определить</u> значения пропускных способностей каналов связи $c_{ij}$ таким образом, чтобы суммарные затраты на организацию сети с выбранными значениями пропускных способностей каналов связи не превышала заранее определенную величину $\Psi_{доп}$.

В качестве <u>критерия оптимальности</u> полученного варианта решения задачи параметрического синтеза является минимум среднесетевой задержки пакета в сети.

### 3. РЕШЕНИЕ ЗАДАЧИ С ИСПОЛЬЗОВАНИЕ МАТЕМАТИЧЕСКОЙ МОДЕЛИ МНОГОСЛОЙНОГО ГРАФА

Базируясь на предметной постановке задачи, приведенной выше, опишем ее математическую постановку.

В качестве математической модели структуры мультисервисной телекоммуникационной системы будем использовать многослойный граф [11]. Многослойный граф $MLG = (\Gamma, V, E)$, включает в свой состав:

- множество подграфов $\Gamma = \{\Gamma^1, ... \Gamma^l, ..., \Gamma^L\}$, $\Gamma^l = (V^l, E^l)$, где подграф $\Gamma^l$ описывает структуру сети на уровне $l$;
- вершины $v_i \in V$ и ребра $e_k = (v_i, v_j)$, $e_k \in E$ обеспечивают связь подграфов $\Gamma^l$ между собой.

На структуру графа MLG, моделирующего мультисервисные телекоммуникационные системы, накладывается дополнительное ограничение, которое заключается в том, что для каждого ребра $e_k^l = (v_i^l, v_j^l)$, $e_k^l \in E^l$ подграфа $\Gamma^l$ существует путь $\pi = (v_i^l, ..., v_m^n, ... v_j^l)$ между вершинами $v_i^l$ и $v_j^l$, $v_i^l, v_j^l \in V^l$, проходящий через подграф более низкого уровня:

$$\forall e_k^l = (v_i^l, v_j^l), \quad e_k^l \in E^l, \quad v_i^l, v_j^l \in V^l, \quad \exists \pi = (v_i^l, ..., v_m^n, ..., v_j^l), v_m^n \in V^n, n < l. \qquad (2)$$

Данное правило не выполняется только для подграфа самого нижнего уровня, $l = 1$.

Пример многослойного графа сети (рис. 2) для случая предоставлении в сети двух услуг приведен на рис. 1.

Согласно общей методике решения задачи синтеза мультисервисной телекоммуникационной системы с использование многослойного графа в структуре синтезируемой сети выделим следующие уровни (слои).

Нижним слоем $l = 1$ многослойного графа $MLG$ является граф $\Gamma^1 = (V^1, E^1)$, описывающий физическую топологию сети.

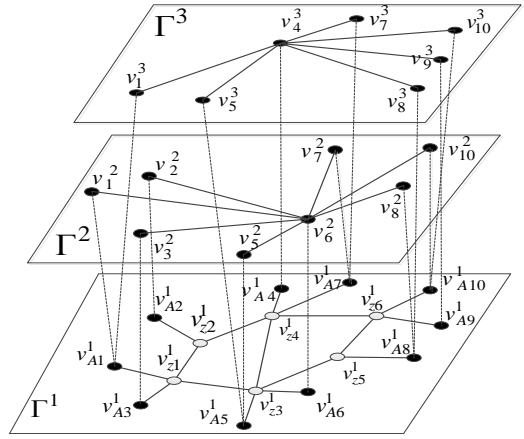

Рисунок 2. — Математическая модель решения задачи параметрического синтеза мультисервисной телекоммуникационной сети.

Слои МСГ $MLG$ выше первого ($l>1$) описывают взаимодействие узлов сети при предоставлении телекоммуникационных услуг. Количество слоев выше первого слоя равно количеству услуг предоставляемых в сети $N_S = |S|$, где $S = \{s_i\}$ - множество услуг предоставляемых сетью. Общее количество слоев графа $L = N_S + 1$.

Вершины графа $\Gamma^1 = (V^1, E^1)$ соответствуют узлам физической сети. Все множество вершин разобьём на два подмножества $V^1 = V_A^1 \bigcup V_Z^1$, где

$V_A^1 = \{v_{Ai}^1\}$ - множество вершин соответствующих узлам сети – источникам (потребителям) информационных потоков;

$V_Z^1 = \{v_{Zi}^1\}$ - множество вершин соответствующих транзитным узлам сети;

$E^1 = \{e_{ij}^1\}$ - множество ребер графа $\Gamma^1$ соответствующих каналам связи физической сети, где $e_{ij}^1 = (v_i^1, v_j^1)$.

Каждому ребру графа $\Gamma^1$ припишем параметр $\alpha_c(e_{ij}^1, c_{ij})$ задающий затраты на организацию канала связи пропускной способностью $c_{ij}$. При решении данной задачи примем затраты имеющие линейную зависимость от пропускной способности канала связи:

$$\alpha_c(e_{ij}^1, c_{ij}^1) = \alpha_{ij} \cdot c_{ij}^1, \qquad (3)$$

где $\alpha_{ij}$ - удельная стоимость единицы пропускной способности канала связи, которому соответствует ребро $e_{ij}^1$.

Слои МСГ $MLG$ выше первого $l>1$ соответствуют услугам, предоставляемым в сети. Вершины $v_i^l \in V^l$ графа $\Gamma^l = (V^l, E^l), l = 2, 3, \ldots, N_S + 1$, соответствуют источникам и потребителям информационных потоков, возникающих при предоставлении телекоммуникационных услуг. Ребра $e_{ij}^l = (v_i^l, v_j^l)$ графа $\Gamma^l$ связывают вершины $v_i^l$ и $v_j^l$, которые соответствуют узлам сети, взаимодействующим друг с другом при предоставлении услуги $s_{l-1} \in S$.

Каждому ребру $e_{ij}^l$ припишем поток $\gamma_{ij}^l \in \Upsilon^l$, где $\Upsilon^l$ - множество информационных потоков передаваемых между узлами сети $a \in A$ при предоставлении услуги $s_{l-1}$. Указанные выше потоки являются групповыми потоками и для их моделирования применяется модель фрактального Броуновского трафика. Зададим параметры передаваемых потоков следующим образом:

$\gamma_{ij}^l = \left(\lambda_{ij}^l, \iota_{ij}^l, \varsigma_{ij}^l, H_{ij}^l\right)$ - набор параметров информационного потока передаваемого между узлом $a_i$ и $a_j$ при предоставлении услуги $s_{l-1}$;

$\lambda_{ij}^l$ - интенсивность потока, бит/с;

$\iota_{ij}^l$ - средняя длина пакета, бит;

$\varsigma_{ij}^l$ - коэффициент дисперсии;

$H_{ij}^l$ - параметр Херста.

Вершины $v_i^l$ графов $\Gamma^l$, $l = 2,\ldots,L$ связаны рёбрами $e_{ij}^{l,1} = \left(v_i^l, v_j^1\right)$ с вершинами $v_j^1$ графа нижнего слоя $\Gamma^1$, которые соответствуют узлам сети $a_j$, где расположен источник или потребитель информационного потока возникающего при предоставлении в сети услуги $s_{l-1}$.

Согласно условию задачи нам известны маршруты передачи потоков в сети между всеми парами $(a_i, a_j)$ узлов отправитель – получатель. Этим маршрутам соответствуют пути $\pi_{\langle i,j \rangle}^1$ протекания потоков между вершинами $v_i^1$ и $v_j^1$ в графе $\Gamma^1$. Путь $\pi_{\langle i,j \rangle}^1$ является упорядоченным множеством рёбер $\pi_{\langle i,j \rangle}^1 = \left(e_{is}^1, \ldots, e_{rj}^1\right)$.

Таким образом, с использованием свойства многослойного графа (2) можно считать путь $\pi_{(km)^l}^1$ соответствующий ребру $e_{km}^l$ верхнего слоя $l > 1$ равным $\pi_{(km)^l}^1 = \left(e_{ki}^{l,1}, e_{is}^1, \ldots, e_{rj}^1, e_{jm}^{1,l}\right)$. Из единства пути $\pi_{\langle i,j \rangle}^1$ (согласно условию задачи) и структуре многослойного графа $MLG$, описывающего синтезируемую мультисервисную сеть, путь $\pi_{(km)^l}^1$ является единственным.

Обозначим как $\gamma_{ij}^1$ - поток, протекающий по ребру $e_{ij}^1$ графа нижнего слоя $\Gamma^1$. Поток $\gamma_{ij}^1$ образуется в результате объединения потоков, соответствующих потокам, протекающих по рёбрам верхних слоёв многослойного графа $MLG$. Таким образом, основываясь на предложенной в работе [12] потоковой модели для потока $\gamma_{ij}^1$ можно записать

$$\gamma_{ij}^1 = \sum_{\substack{l=2,\ldots,L, \\ e_{km}^l \in E^l, \, e_{ij}^1 \in \pi_{(km)^l}^1}} \gamma_{km}^l . \qquad (4)$$

Параметры агрегированных потоков протекающих по рёбрам графа $\Gamma^1$ можно определить согласно методики, описанной в работе [13]. На основании этого можно записать

$$\lambda_{ij}^1 = \sum_{\substack{l=2,\ldots,L, \\ e_{km}^l \in E^l, \, e_{ij}^1 \in \pi_{(km)^l}^1}} \lambda_{km}^l , \qquad (5)$$

$$\iota_{ij}^1 = \sum_{\substack{l=2,\ldots,L, \\ e_{km}^l \in E^l, \, e_{ij}^1 \in \pi_{(km)^l}^1}} \left(\iota_{km}^l \lambda_{km}^l\right) \Bigg/ \left[\sum_{\substack{l=2,\ldots,L, \\ e_{km}^l \in E^l, \, e_{ij}^1 \in \pi_{(km)^l}^1}} \lambda_{km}^l\right], \qquad (6)$$

$$\varsigma_{ij}^1 = \sum_{\substack{l=2,\ldots,L, \\ e_{km}^l \in E^l, e_{ij}^1 \in \pi_{(km)^l}^1}} \left(\varsigma_{km}^l \lambda_{km}^l\right) \Bigg/ \left[\sum_{\substack{l=2,\ldots,L, \\ e_{km}^l \in E^l, e_{ij}^1 \in \pi_{(km)^l}^1}} \lambda_{km}^l\right], \quad (7)$$

$$H_{ij}^1 = \max_{\substack{l=2,\ldots,L, \\ e_{km}^l \in E^l, e_{ij}^1 \in \pi_{(km)^l}^1}} \left[H_{km}^l\right]. \quad (8)$$

В этом случае среднесетевую задержку, с использованием выражения, полученного Норросом, для средней длины очереди [4] согласно формулы Литтла [14] можно определить как

$$\overline{T}\left(\Gamma^1, \Upsilon, c(e^1)\right) = \frac{1}{\Lambda} \sum_{e_{ij}^1 \in E^1} \left[\frac{\lambda_{ij}^1}{c_{ij}^1}\left(1 + \frac{\left(\lambda_{ij}^1\right)^{(2H_{ij}^1-1)/(2-2H_{ij}^1)} \cdot c^{1/(2-2H_{ij}^1)}}{\left(c_{ij}^1 - \lambda_{ij}^1\right)^{H_{ij}^1/(1-H_{ij}^1)}}\right)\right], \quad \Upsilon = \left\{\Upsilon^l\right\}, \quad (9)$$

$$\Lambda = \sum_{l=2,\ldots,L} \sum_{\substack{\gamma_{km}^l, \\ e_{km}^l \in E^l}} \lambda_{km}^l \Big/ \mathfrak{l}_{km}^l. \quad (10)$$

Таким образом, основываясь на приведенной математической модели мультисервисной телекоммуникационной системы, задачу параметрического синтеза сформулируем как оптимизационную задачу следующего вида.

<u>Задано:</u>
$MLG = \left(\left\{\Gamma^l\right\}, E, \alpha(e^1)\right)$ - многослойный граф, описывающий структуру мультисервисной телекоммуникационной системы;

$\Upsilon = \left\{\Upsilon^l\right\}$ - множество потоков протекающих по ребрам многослойного графа $MLG$.

<u>Найти:</u>
$c_{ij}^1, \forall e_{ij}^1 \in E^1$ - пропускные способности ребер графа $\Gamma^1$ соответствующих каналам связи синтезируемой мультисервисной телекоммуникационной сети.

<u>Критерий оптимальности:</u>

$$\overline{T}\left(\Gamma^1, \Upsilon, c(e^1)\right) \to \infty. \quad (11)$$

<u>Ограничения:</u>

$$\sum_{e_{ij}^1 \in E^1} \alpha_{ij} \cdot c_{ij}^1 \leq \Psi_{\text{доп}}, \quad (12)$$

$$\gamma_{ij}^1 = \sum_{\substack{l=2,\ldots,L, \\ e_{km}^l \in E^l, e_{ij}^1 \in \pi_{(km)^l}^l}} \gamma_{km}^l \quad (13)$$

$$\lambda_{ij}^1 < c_{ij}^1, \quad \forall e_{ij}^1 \in E^1, \quad (14)$$

Выражение (11) является целевой функцией и формирует критерий минимума среднесетевой задержки пакета в сети. Значение функции $\overline{T}\left(\Gamma^1, \Upsilon, c(e^1)\right)$ определяется из выражения (9), (10).

Условие (12) является ограничением на максимально допустимые расходы на организацию каналов связи, искомой пропускной способностью.

Выражение (13) позволяет определить параметры результирующего потока в каналах связи синтезируемой сети. При этом частные параметры агрегированного потока необходимо определять из (5) - (8).

Условие (14) гарантирует не превышение интенсивности потока -значения пропускной способности ребер графа (пропускной способности каналов связи).

Анализ целевой функции (11) показал, что она является строго выпуклой функцией и следовательно, существует единственная стационарная точка, которая и является глобальным минимумом.

С другой стороны анализ целевой функции и ограничений задачи показал, что решение задачи лежит на границе допустимой области и в этом случае неравенство (12) превращается в равенство. Это позволяет с применением условия (12) в виде равенства выразить пропускную способность одного из ребер через пропускные способности других ребер. После подстановки найденной зависимости в целевую функцию задача с ограничениями преобразуется в задачу без ограничений, что позволяет напрямую применить метод наискорейшего спуска.

Проведенные эксперименты показали работоспособность и эффективность предлагаемого метода параметрического синтеза. Сопоставление результатов, получаемых при применении предложенного в работе метода параметрического синтеза при значении параметра Херста равного $H = 0,5$ и ранее известного метода базирующегося на моделях простейшего потока, показало сходимость результатов расчетов. Это позволяет сделать вывод об общности предлагаемых моделей и методов параметрического синтеза

Сопоставление результатов имитационного моделирования двух конфигураций сети, рассчитанных с применением предлагаемого авторами метода и классического метода «квадратного корня» показало, что предлагаемый метод позволяет более точно определять параметры каналов связи и обеспечивает уменьшение средней задержки от 10 до 25%.

## ЗАКЛЮЧЕНИЕ

1. При синтезе мультисервисных телекоммуникационных сетей, одной из наиболее важных задач является обеспечение требуемого качества обслуживания, которое достигается при параметрическом синтезе. Методы параметрического синтеза позволяют использовать, как линейных, так и нелинейных модели и учитывать наличие функциональных и вероятностных зависимостей между параметрами передаваемых потоков и параметрами качества обслуживания.

2. Для разработки методики синтеза мультисервисных сетей в качестве математических моделей потоков на различных уровнях и участках мультисервисной наложенной телекоммуникационной сети целесообразно использовать модели самоподобных процессов. Указанные модели позволяют учитывать такие свойства характерные потокам в мультисервисных сетях как: долговременную зависимость, высокую пачечность, наличие распределения с тяжелыми хвостами для межпакетных интервалов и длительностей занятий обслуживающих устройств, медленно затухающая дисперсия выборочного среднего.

3. Математические модели потоков, как самоподобных процессов, и полученные расчетные выражения были использованы при разработке методики параметрического синтеза мультисервисных телекоммуникационных сетей, которая позволила решить задачу выбора пропускных способностей каналов связи при использовании критерия минимума среднесетевой задержки и ограничения на величину максимально допустимых затрат на организацию сети.

4. Применение предлагаемого в статье метода параметрического синтеза позволяет более эффективно распределить пропускные способности каналов связи и уменьшить на 10 – 25 % среднесетевую задержку чем ранее известный метод базирующийся на моделях простейшего потока.

5. В качестве математической модели структуры мультисервисной телекоммуникационной сети использовался многослойный граф, нижний слой которого описывал топологию синтезируемой сети, а верхние слои соответствовали услугам, предоставляемым в сети. Разработанные методики доказывают возможность применения, приведенных в разделе математических моделей и расчетных выражений, для решения задач параметрического синтеза мультисервисных телекоммуникационных сетей, при решении задачи для всей сети в целом.

6. Разработанный метод параметрического синтеза используются в процессе проектирования, и являются обоснованием для выбора значений конфигурационных параметров телекоммуникационного оборудования в узлах сети при планировании и эксплуатации мультисервисных телекоммуникационных систем, а также составлении спецификаций, устанавливаемого оборудования, при развертывании новых и модернизации существующих мультисервисных телекоммуникационных систем.

## СПИСОК ИСПОЛЬЗОВАННЫХ ИСТОЧНИКОВ